\documentclass[pra,showkeys,twocolumn,showpacs]{revtex4}
\usepackage{amsmath}
\usepackage{amsfonts}
\usepackage{amssymb}
\usepackage{array}
\usepackage{floatflt}
\usepackage{graphicx}
\DeclareMathOperator{\tr}{Tr}
\newcommand*{\eq}[1]{Eq.~(\ref{#1})}

\begin{document}
\title{The effect of spin-lattice relaxation on multiple-quantum NMR coherences in solids}
\author{E.~B.~Fel'dman, A.~N.~Pyrkov\footnote{Email address:
pyrkov@icp.ac.ru}} \affiliation{Institute of Problems of Chemical
Physics of Russian Academy of Sciences, Chernogolovka, Moscow
Region, Russia, 142432}

\date{\today}

\begin{abstract}
A phenomenological theory of spin-lattice relaxation of multiple-quantum coherences in systems of two dipolar coupled spins at low temperatures is developed. Intensities of multiple-quantum NMR coherences depending on the spin-lattice relaxation time are obtained. It is shown that the theory is also applicable to finite spin chains when the approximation of nearest neighbour interaction is used. An application of this theory to an estimation of the influence of decoherence processes on quantum entanglement and its fluctuations is briefly discussed.
\end{abstract}

\maketitle
\section{Introduction}
Multiple-quantum (MQ) coherences in a system of dipolar coupled spins can be detected by MQ NMR methods~\cite{pines,cho}. In the simplest cases (one-dimensional spin chains and rings, a system of spin-carrying atoms (molecules) in nanopores), analytical and numerical methods were developed~\cite{lacelle, nanopore} in order to calculate the intensities of MQ NMR coherences. Usually spin-lattice relaxation is not taken into account at investigations of MQ NMR dynamics because the characteristic times of spin-lattice relaxation are longer than the characteristic time of MQ NMR dynamics~\cite{cho}. At the same time, MQ NMR methods have been also used to measure decoherence for highly correlated spin states~\cite{suter, cory} and to study the scaling of the decoherence rate with the number of correlated spins~\cite{suter}. MQ NMR investigations showed both experimentally~\cite{alvarez} and with computer simulations for a system of equivalent spins~\cite{doronin} that the relaxation time decreases with the increase in the MQ coherence order and in the number of spins. It was also shown~\cite{jetpl_2008} that MQ NMR allows us to find a relationship between the intensity of the MQ coherence of the second order and a measure of the quantum entanglement which is the main resource of quantum computers~\cite{quant_comp}. However, that relationship was found without taking into account effects of spin-lattice relaxation~\cite{jetpl_2008}.  A theoretical interpretation of such results is impossible without taking into account those effects. A theory of spin-lattice relaxation is also useful for traditional problems of NMR (paramagnetic impurities, internal molecular motions, etc.)~\cite{abragam}.

In the present paper we develop a phenomenological theory of spin-lattice relaxation for a system consisting of two dipolar coupled spins in the MQ NMR experiment at low temperatures. We calculate the intensities of MQ coherences of the zeroth and plus/minus second orders and discuss the effect of spin-lattice relaxation on MQ NMR dynamics. We show that the developed approach is also applicable for linear spin chains when the approximation of nearest neighbor interactions is used. We discuss a connection of quantum entanglement with the intensity of MQ NMR coherence of the second order at different spin-lattice relaxation times.

\section{MQ NMR of spin pairs with spin-lattice relaxation at low temperatures}
We consider a system consisting of two dipolar coupled spins ($s=1/2$) in the conditions of the MQ NMR experiment~\cite{pines}. Initially, the system is in the thermodynamic equilibrium state in a strong external magnetic field $\overrightarrow{H}_0.$ The density matrix $\rho_{eq}$ of this state in the laboratory reference frame is given by
\begin{equation}
\label{rho0} \rho_{eq}=\frac{\exp(\frac{\hbar \omega_0}{kT}I_z)}{\tr{\{\exp(\frac{\hbar \omega_0}{kT}I_z)\}}}=\frac{\exp(\frac{\hbar \omega_0}{kT}I_z)}{4\cosh^2(\frac{\hbar \omega_0}{2kT})},
\end{equation}
where $\omega_0=\gamma\left|\overrightarrow{H}_0\right|$ is the Larmor frequency,$\gamma$ the gyromagnetic ratio, $T$ the temperature, $k$ the Bolzman constant, $I_z=I_{z1}+I_{z2}$, and $I_{\alpha i} (\alpha=x,y,z; i=1,2)$ is the operator of the projection of spin $i$ onto the axis $\alpha.$

We can neglect indirect spin--spin interactions and chemical shifts in~\eq{rho0} because they are much smaller than dipole--dipole interactions (DDI). The Hamiltonian of the DDI of the two spins, $H_{dz},$ in the strong external magnetic field~\cite{abragam} is
\begin{equation}
\label{hdz}
H_{dz}=D(3I_{z1}I_{z2}-\vec{I}_{1}\vec{I}_{2}),
\end{equation}
where $D$ is the coupling constant and $\vec{I}_{1}\vec{I}_{2}=I_{x1}I_{x2}+I_{y1}I_{y2}+I_{z1}I_{z2}.$

We study MQ NMR dynamics of this two--spin system. The MQ NMR experiment~\cite{pines} consists of four periods: preparation, evolution, mixing and detection. The spin system is irradiated on the preparation period by the multi--pulse sequence in which every period consists of eight resonance pulses~\cite{pines}. It is suitable to switch to the rotating (with the Larmor frequency) reference frame~\cite{abragam} where the multi--pulse sequence consists of the four bloks of the form $\bar{\tau}/2-90^{\circ}_x-2\bar{\tau}-90^{\circ}_{-x}-\bar{\tau}/2$ where $\bar{\tau}$ and $2\bar{\tau}$ are the time intervals between consecutive pulses and $90^{\circ}_{\pm x}$ denotes pulses turning spins about axis $\pm x$ \cite{pines}. The DDI are oscillating fast, if the period $4*3\bar{\tau}=12\bar{\tau}$ of the multi--pulse sequence is less than the inverse coupling constant $D^{-1}$ \cite{waugh}. As a result, spin dynamics of the system on the preparation period of the MQ NMR experiment is governed by the averaged non-secular Hamiltonian~\cite{pines,waugh} 

\begin{equation}
\label{HMQ} H_{MQ}=-\frac{D}2\left( I_1^{+}I_2^{+}+I_1^{-}I_2^{-}\right),
\end{equation}
where $I_i^+$ and $I_i^-$ ($i=1,2$) are raising and lowering operators ($I_i^{\pm}=I_{xi}\pm iI_{yi}, i=1,2$). Spin dynamics on the preparation period is described by the Liouville equation for the density matrix $\rho(t)$~\cite{abragam},
\begin{equation}
\label{liouv}
\imath\frac{d\rho(t)}{dt}=[H_{MQ},\rho(t)],
\end{equation}
and $\rho(0)=\rho_{eq}.$ It is worth to emphasize that MQ NMR coherences are determined by the density matrix at the end of the preparation period of the MQ NMR experiment~\cite{pines,lacelle}. 

In order to take into account spin-lattice relaxation one has to introduce, generally speaking, a non--linear term with respect to the density matrix $\rho(t)$ in the right hand side of~\eq{liouv}. In practice, such equation can be solved with the linearization of the right hand side after it is expressed as an expansion in powers of $\rho(t)-\overline{\rho}$, where $\overline{\rho}$ is the equilibrium density matrix in the RRF. This procedure is valid, if the deviation of $\rho(t)$ from $\overline{\rho}$ is small. One can expect a small deviation, if the time independent $\rho=\overline{\rho}$ is a solution of the Liouville equation. It is reasonable to suppose that~\cite{goldman1}
\begin{equation}
\label{rhoeqpr} \overline{\rho}=\frac{\exp(-\frac{\hbar H_{MQ}}{kT})}{\tr{\{\exp(-\frac{\hbar H_{MQ}}{kT})\}}}.
\end{equation}
The temperature T is of an order of the lattice temperature. Its exact value is not significant in our consideration.

We will consider the millikelvin temperature range, where one can suppose~\cite{goldman} that $\frac{\hbar D}{kT}\ll1,$ $\overline{\rho}=1/4e$ and $e$ is the 4 by 4 unit matrix.
 Then the Liouville equation with a linearized phenomenological term for spin--lattice relaxation reads
\begin{equation}
\label{liouv1}
\frac{d\rho(t)}{dt}=-\imath[H_{MQ},\rho(t)]-\frac{\rho(t)-\overline{\rho}}{T_{MQ}},
\end{equation}
where $T_{MQ}$ is the spin--lattice relaxation time of MQ NMR coherences. A reorienting spin group can be responsible for relaxation~\cite{zobov}. We consider relaxation only as a pertubation of MQ NMR dynamics, i.e. we mean that $D\gg 1/T_{MQ}.$ We assume also that the spin--lattice relaxation time $T_1$ in the laboratory reference frame is large $(T_1\gg T_{MQ})$ and we neglect it. It is worth to emphasize that the equilibrium density matrix $\overline{\rho}$ describes the spin system in the RRF, where there is no any external magnetic field, i.e. the Zeeman reservoir is absent in the considered case. Thus, relaxation of MQ NMR coherences can be described with only one relaxation time, $T_{MQ}.$ Introducing $\rho^*(t)=\rho(t)-\overline{\rho}$ and $\overline{\rho}^*(t)=\rho^*(t)e^{t/T_{MQ}}$ one can find that~\eq{liouv1} can be written as
\begin{equation}
\label{liouv2}
\frac{d\overline{\rho}^*(t)}{dt}=-\imath[H_{MQ},\overline{\rho}^*(t)],
\end{equation}
and $\overline{\rho}^*(0)=\rho(0)-\overline{\rho}=\rho_{eq}-\overline{\rho}.$ At the end of the preparation period $\tau$ of the MQ NMR experiment, the solution of~\eq{liouv2} can be written  as
\begin{equation}
\label{solv1}
\overline{\rho}^*(\tau)=e^{-\imath H_{MQ}\tau}\rho_{eq}e^{\imath H_{MQ}\tau}-\overline{\rho}.
\end{equation}
The first term in the right hand side of~\eq{solv1} was obtained in~\cite{jetpl_2008},
\begin{multline}
\label{term1}
e^{-\imath H_{MQ}\tau}\rho_{eq}e^{\imath H_{MQ}\tau}=\frac1{4\cosh^2\beta/2}\times\\
\left(\begin{smallmatrix}
\cosh\beta+\cos(D\tau)\sinh\beta&0&0&-i\sin(D\tau)\sinh\beta\\
0&1&0&0\\
0&0&1&0\\
i\sin(D\tau)\sinh\beta&0&0&\cosh\beta-\cos(D\tau)\sinh\beta
\end{smallmatrix}\right)
\end{multline}
where $\beta=\frac{\hbar\omega_0}{kT}.$ Using~\eq{solv1} one can find that
\begin{equation}
\label{solv2}
\rho(\tau)=e^{-\imath H_{MQ}\tau}\rho_{eq}e^{\imath H_{MQ}\tau}e^{-\tau/T_{MQ}}+(1-e^{-\tau/T_{MQ}})\overline{\rho}.
\end{equation}
The density matrix $\rho(\tau+t)$ after the evolution period $t$ of the MQ NMR experiment~\cite{pines}, during which the decoding field $\Delta$ (in frequency units) acts, is
\begin{multline}
\label{solv_ev}
\rho(\tau+t)=e^{-\imath\Delta tI_z}e^{-\imath H_{MQ}\tau}\rho_{eq}e^{\imath H_{MQ}\tau}e^{\imath\Delta tI_z}e^{-\tau/T_{MQ}}+\\(1-e^{-\tau/T_{MQ}})\overline{\rho}.
\end{multline}
We did not consider spin-lattice relaxation on the evolution period since we assumed that $t\ll T_{MQ}.$ Finally, after the mixing period $\tau,$ during which the Hamiltonian $H_{MQ}$ changes its sign, the density matrix $\rho(2\tau+t)$ becomes
\begin{multline}
\label{solv_exp}
\rho(2\tau+t)=\\e^{\imath H_{MQ}\tau}e^{-\imath\Delta tI_z}e^{-\imath H_{MQ}\tau}\rho_{eq}e^{\imath H_{MQ}\tau}e^{\imath\Delta tI_z}e^{-\imath H_{MQ}\tau}e^{-2\tau/T_{MQ}}+\\(1-e^{-2\tau/T_{MQ}})\overline{\rho}.
\end{multline}
Calculating the longitudinal magnetization $I_z(2\tau+t)$ at the end of the MQ NMR experiment~\cite{pines} and taking into account that $\tr(I_z\overline{\rho})=0$ one can find that only MQ NMR coherences of zero and plus/minus second orders emerge in the our system and their intensities $J_0(\tau),$ $J_{\pm2}(\tau),$ are:
\begin{multline}
\label{cohs} J_0(\tau)=\tanh\frac{\beta}2\cos^2(D\tau)e^{-2\tau/T_{MQ}},\\
J_{\pm2}(\tau)=\frac12\tanh\frac{\beta}2\sin^2(D\tau)e^{-2\tau/T_{MQ}}.
\end{multline}
Note that all performed calculations  are also valid for any equilibrium density matrix $\overline{\rho}$ which commutes with operators $H_{MQ}$ and $I_z.$ 

It is also interesting to note that an analogous approach can be applied to a linear chain of $N$ interacting spins in the approximation of nearest neighbour interactions~\cite{maximov}. Only MQ NMR coherences of zero and plus/minus second orders emerge in such system. Using exact expressions for the intensities of these MQ NMR coherences~\cite{maximov}, which were obtained without taking into account spin-lattice relaxation, one can obtain
\begin{multline}
\label{cohs_chain} J_0(\tau)=\frac{e^{-2\tau/T_{MQ}}}{N}\sum_k\cos^2(2D\tau\cos k),\\
J_{\pm2}(\tau)=\frac{e^{-2\tau/T_{MQ}}}{N}\sum_k\sin^2(2D\tau\cos k),
\end{multline}
where $k=\frac{\pi n}{N+1},\quad(n=1,2,\ldots,N).$ It is worth to emphasize that we neglect dipole-dipole interactions of next nearest neighbours which are only eight times weaker than the interactions of nearest neighbours. It means that the developed approach is justified only when
\begin{equation}
D\gg1/T_{MQ}\gg D/8.
\end{equation}
In considered systems we have found a simple exponential law for spin--lattice relaxation of MQ NMR coherences.

\section{Quantum entangled states in MQ NMR}
It is well known~\cite{quant_comp} that entanglement is the main resource for quantum devices (in particular, quantum computers). MQ NMR is one of the numerous methods of obtaining entangled states~\cite{jetpl_2008,furman}. It turned out~\cite{jetpl_2008} that intensity of the MQ NMR coherence of the second order (without spin-lattice relaxation) is intimately connected with the concurrence~\cite{woot} which describes quantitatively entanglement of quantum systems. In order to calculate the concurrence~\cite{woot} at the end of the preparation period one should determine the matrix $\tilde{\rho}(\tau)$ as
\begin{equation}
\label{flip}
\tilde{\rho}(\tau)=(\sigma_{y1}\otimes\sigma_{y2})\rho_{cc}(\tau)(\sigma_{y1}\otimes\sigma_{y2})
\end{equation}
where $\rho_{cc}(\tau)$ is the complex conjugate matrix $\rho(\tau)$ of \eq{solv2} in the standard basis
$\{|00\rangle,|01\rangle,|10\rangle,|11\rangle\}$ and $\sigma_{yi}=2I_{yi},$ $(i=1,2)$ is the Pauli
matrix.

The concurrence $C$ of the two-spin system with the density matrix $\rho(\tau)$ of~\eq{solv2} equals to~\cite{woot}
\begin{multline}
\label{conc} C=max\{0,
2\lambda-\lambda_1-\lambda_2-\lambda_3-\lambda_4\},\\
\lambda=max\{\lambda_1,\lambda_2, \lambda_3, \lambda_4\}
\end{multline}
where $\lambda_1,$ $\lambda_2,$ $\lambda_3,$ and $\lambda_4$ are the
square roots of the eigenvalues of the matrix product
$\rho(\tau)\tilde{\rho}(\tau).$ Although this product is not Hermitian, it has real nonnegative eigenvalues~\cite{horn}. One can find from~Eqs.~(\ref{term1}),(\ref{solv2}),(\ref{flip}),(\ref{conc}) that the concurrence $C(\tau)$ in our system is
\begin{multline}
\label{concdec}
C(\tau)=max\left[\right.0,\frac{e^{-\tau/T_{MQ}}}{2\cosh^2\beta/2}(|\sin(D\tau)|\sinh\beta-1)-\\\frac{1-e^{-\tau/T_{MQ}}}{2}\left.\right].
\end{multline}
This expression coincides with the one obtained in~\cite{jetpl_2008} in the limit of the infinite spin--lattice relaxation time. The entangled state emerges at temperatures
\begin{equation}
\label{temp} T<T_E=\frac{\hbar\omega_0}{k\ln(\frac{\sqrt{2+a}}{2-\sqrt{2+a}})},
\end{equation}
where $a=e^{\tau/T_{MQ}}-1.$ If one takes $\omega_0=2\pi 500\cdot10^6 s^{-1}$ the entangled state
emerges at the temperature $T_E\approx27mK$ when $T_{MQ}\rightarrow\infty$~\cite{jetpl_2008}. According to \eq{temp} the entangled states emerge at $2>e^{\tau/T_{MQ}}-1,$ i.e. at $\ln3>\tau/T_{MQ}.$ The temperature $T_E$ decreases with the decrease in  the spin-lattice relaxation time.  Using~Eqs.~(\ref{cohs}) and (\ref{concdec}) one can find a simple relationship between intensities of MQ NMR coherences of the second order and the concurrence:
\begin{multline}
\label{cc}
C=\sqrt{\tanh\frac{\beta}2[J_2(\tau)+J_{-2}(\tau)]}-\\\frac{e^{-\tau/T_{MQ}}}{2\cosh^2\beta/2}-\frac{1-e^{-\tau/T_{MQ}}}{2}.
\end{multline}
This formula coincides again with the analogous one obtained in~\cite{jetpl_2008} at $T_{MQ}\rightarrow\infty.$

In order to obtain the entanglement measure of the two-qubit system we average over the state of the first spin and obtain the reduced density matrix for the second spin~\cite{woot}. The reduced density matrix thus depends on the fluctuations of the first spin. These fluctuations give rise to the fluctuations of the entanglement measure~\cite{fluct}.

The rms fluctuations, $\Delta E,$ of the entanglement entropy of the two-qubit system are~\cite{fluct}
\begin{equation}
\Delta E=C\log_2[\frac1{C}(1+\sqrt{1-C^2})],
\end{equation}
where the concurrence $C$ is determined by \eq{concdec}.

\begin{figure}
\includegraphics[width=7.5cm]{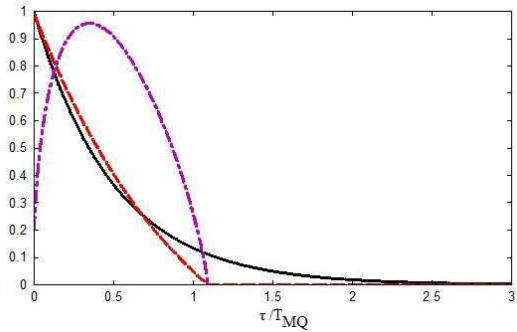}
\caption{The sum of the MQ NMR coherences of the plus and minus second orders (solid line), concurrence $C(\tau)$ (dash line), the entanglement fluctuations (dash-dot line) versus the dimensionless spin-lattice relaxation rate $\tau/T_{MQ}.$  The spin-spin coupling constant $D=4\pi1307 s^{-1},$ the duration of the preparation period $\tau=0.00086 s,$ $D\tau=9\pi/2,$ and $\beta=6.$} \label{fig}
\end{figure}

The dependencies of the sum of the MQ NMR coherences of the plus/minus second order, the concurrence $C(\tau)$ and the entanglement fluctuations on the dimensionless spin-lattice relaxation rate $\tau/T_{MQ}$ at $\beta=6$ are shown in~Fig.~\ref{fig}. At long spin-lattice relaxation times the concurrence approximately coincides with the sum of the intensities of the MQ NMR coherences of the plus/minus second order as it was obtained in~\cite{jetpl_2008}. At $DT_{MQ}\gg1,$ MQ NMR dynamics determines the entanglement and its fluctuations. Spin--lattice relaxation is the dominant factor when $DT_{MQ}\approx1.$ We note also that the concurrence decreases with a decrease of the spin-lattice relaxation time and entanglement fluctuations are getting more important.

In conclusion we emphasize that quantum entanglement can be studied with MQ NMR methods and the effect of spin-lattice relaxation on MQ NMR coherences is important both for traditional magnetic resonance~\cite{abragam} and quantum information processing~\cite{quant_comp}.

\section{Acknowledgments}

We are grateful to D.E. Feldman for many helpful discussions. This work was supported by the Program of the Presidium of Russian Academy of Sciences No. 7 (the program "Development of the methods for obtaining chemical substances and creating new materials").

\end{document}